\documentclass[11pt,draftcls,peerreview,onecolumn]{IEEEtran}

\usepackage[dvips]{graphics}
\usepackage{graphicx}
\usepackage{times}
\usepackage{amsmath}
\usepackage{amssymb}
\usepackage{fancybox}
\newtheorem{lemma}{\bf Lemma}
\newtheorem{theorem}{\bf Theorem}

\newcommand{\bee}{\begin{eqnarray}}
\newcommand{\eee}{\end{eqnarray}}
\newcommand{\be}{\begin{equation}}
\newcommand{\ee}{\end{equation}}

\newcommand{\al}[1]{\begin{align} #1 \end{align}}

\newcommand{\equ}[1]{\begin{equation} #1 \end{equation}}

\newcommand{\mb}{\mathbf}
\newcommand{\mr}{\mathrm}
\newcommand{\bs}{\boldsymbol}

\newcommand{\nnb}{\nonumber}

\newcommand{\qa}{{\bf a}}

\newcommand{\qh}{{\bf h}}

\newcommand{\qu}{{\bf u}}

\newcommand{\qx}{{\bf x}}

\newcommand{\qz}{{\bf z}}

\newcommand{\qA}{{\bf A}}
\newcommand{\qB}{{\bf B}}

\newcommand{\qD}{{\bf D}}

\newcommand{\qI}{{\bf I}}
\newcommand{\qJ}{{\bf J}}
\newcommand{\qK}{{\bf K}}

\newcommand{\qQ}{{\bf Q}}
\newcommand{\qR}{{\bf R}}

\newcommand{\qU}{{\bf U}}
\newcommand{\qV}{{\bf V}}

\newcommand{\qX}{{\bf X}}
\newcommand{\qY}{{\bf Y}}
\newcommand{\qZ}{{\bf Z}}

\begin{document}
%
% paper title
% can use linebreaks \\ within to get better formatting as desired
\title{On Ergodic Secrecy Capacity for Gaussian MISO Wiretap Channels}

% author names and affiliations
% use a multiple column layout for up to three different
% affiliations
\author{\IEEEauthorblockN{Jiangyuan Li and Athina P. Petropulu}\\
\IEEEauthorblockA{Department of Electrical and Computer Engineering\\
Drexel University, Philadelphia, PA 19104}\\
\IEEEauthorblockA{Email: eejyli@yahoo.com.cn, athina@coe.drexel.edu}
}

% make the title area
\maketitle

\begin{abstract}
\footnote{Work supported by the Office of Naval Research under grant ONR-N-00010710500 and the
National Science Foundation under grant CNS-0905425.}
A Gaussian multiple-input single-output (MISO) wiretap channel model is considered,
where there exists a transmitter equipped with multiple antennas, a legitimate receiver and an eavesdropper each equipped with
a single antenna.
We study the problem of finding the optimal input covariance that achieves ergodic secrecy capacity subject to a power constraint
where only statistical information about the eavesdropper channel is available at the transmitter.
This is a non-convex optimization problem that is in general difficult to solve.
Existing results address the case in which the eavesdropper or/and legitimate channels
have independent and identically distributed Gaussian entries with zero-mean and unit-variance, i.e., the channels have trivial covariances.
This paper addresses the general case where  eavesdropper and legitimate channels have  nontrivial covariances.
A set of equations describing the optimal input covariance matrix are proposed along with an algorithm to obtain
the solution.
Based on this framework,
we show that when full  information on the legitimate channel is available to the transmitter,
the optimal input covariance has always rank one. We also show that when only statistical information on the legitimate channel
is available to the transmitter, the legitimate channel has some general non-trivial covariance, and the eavesdropper channel has trivial covariance,
the optimal input covariance has the same eigenvectors as the legitimate channel covariance.
Numerical results are presented to illustrate the algorithm.
\end{abstract}


%\IEEEpeerreviewmaketitle
\newpage

\section{Introduction}

Wireless physical (PHY) layer based security from a information-theoretic point
of view has received considerable attention recently \cite{Liang2}.
Such approaches exploit the physical characteristics of the
wireless channel to enhance the security of communication systems.
The wiretap channel, first introduced and studied by Wyner \cite{Wyner}, is the most basic physical layer model that captures
the problem of communication security.
Wyner showed that when an eavesdropper's channel is a degraded version of the legitimate channel,
the source and destination can achieve a positive information rate
(secrecy rate).
The maximal secrecy rate from the source to the destination
is defined as the {\em secrecy capacity};  for the degraded wiretap channel the secrecy capacity  is given as the
 largest between  zero and
the difference between the capacity at the legitimate receiver and the capacity at the eavesdropper.
The Gaussian wiretap channel, in which the outputs at the legitimate
receiver and at the eavesdropper are corrupted by additive white
Gaussian noise (AWGN), was studied in \cite{Hellman}.
Along the same lines, the secrecy capacity
of a deterministic Gaussian MIMO wiretap channel has been studied recently
in \cite{Khisti}-\cite{Shafiee}. In \cite{Shafiee2}, the achievable rate in Gaussian MISO channels was studied.
In that context,  the channel state information (CSI) of the legitimate channel was assumed to be available, but only statistical information
about the eavesdropper channel was assumed to be available at the transmitter. In \cite{Shafiee2} it was shown that when the eavesdropper channel
is a vector of independent and identically distributed (i.i.d.) zero-mean complex circularly symmetric Gaussian
random variables, i.e., the channel has a trivial covariance matrix, the optimal communication strategy is beamforming, and that the beamforming direction depends on
the CSI of the legitimate channel.
In \cite{Bhargava}, the authors derived the ergodic secrecy capacity of a
Gaussian MIMO wiretap channel where only  statistical information about the legitimate and eavesdropper channels are available at the transmitter. It was shown that a circularly symmetric Gaussian input is optimal.
It was  also shown in the same paper that when the eavesdropper and legitimate channels
have i.i.d. Gaussian entries with zero-mean and unit-variance (trivial covariance),
a circularly symmetric Gaussian input with diagonal covariance is optimal.

In this paper, we consider a  Gaussian multiple-input single-output (MISO) wiretap channel
and assume that only statistical information about the  the eavesdropper channel is available at the transmitter.
Regarding the legitimate channel, we  consider two scenarios:
a) only  statistical information of the legitimate  channel is available at the transmitter;
b) full CSI on the legitimate channel is available at the transmitter.
We extend the result of \cite{Shafiee2} and \cite{Bhargava} proposed for the case of multiple-input single-output (MISO)
wiretap channel with trivial channel covariances to the case of nontrivial covariances.
The non-trivial channel covariance matrix
corresponds to the case where there exists statistical correlation between the channel coefficients of different transmit-receive
antenna pairs. Such cases arise when the transmit and receive antennas are closely spaced relative to the signal wavelength.
We address the problem of finding the optimal input covariance that achieves ergodic secrecy capacity subject to a power constraint.
This leads to a non-convex optimization problem.
The contributions of this paper are the following:
\begin{itemize}
\item We derive a set of equations for the optimal input covariance matrix,  and
propose an algorithm to obtain the solution (please refer to Theorem 1 of Section IV).

\item We show that when the legitimate channel is completely known at the transmitter,  in addition to the  conditions of Theorem \ref{theo:1},
the following hold:
1) the optimal input covariance matrix has rank one; 2) the ergodic secrecy rate is increasing with the signal-to-noise ratio (SNR).

\item We show that when only statistical information on the legitimate channel
is available to the transmitter, the legitimate channel has some general non-trivial covariance, and the eavesdropper channel has trivial covariance,
the optimal input covariance has the same eigenvectors as the legitimate channel covariance.

\item We show that under high SNR, the optimal input covariance has rank one.
\end{itemize}

The remainder of this paper is organized as follows.
The mathematical model is introduced in \S\ref{sec:2}.
In \S\ref{sec:CsQ}, we give the explicit expression of ergodic secrecy rate,
and in \S\ref{OptInpCov}, we derive the condition for optimal input covariance.
In \S\ref{subSec:ImpactOnSNR}, we analyze the dependence of ergodic secrecy rate on the SNR,
and in \S\ref{Sec:HighSNR}, we study the ergodic secrecy rate under high SNR.
In \S\ref{Sec:Algo}, an algorithm is proposed to search for the solution.
Numerical results are presented in \S\ref{sec:sim} to illustrate the proposed algorithm.
Finally, \S\ref{sec:conclusion} gives a brief conclusion.  Several proofs appear in an Appendix.

\subsection{Notation}

Upper case and lower case bold symbols denote matrices and vectors, respectively.
Superscripts $\ast$, $T$ and $\dagger$
denote respectively conjugate, transposition and conjugate transposition. $\mathrm{det}(\qA)$ and
$\mathrm{Tr}({\mb A})$ denote the determinant and trace of matrix $\mb A$, respectively.
$\lambda_{\max}(\qA)$ and $\lambda_{\min}(\qA)$ denote the largest and smallest eigenvalues of $\qA$, respectively.
${\mb A}\succeq 0$ means that $\qA$ is Hermitian positive semi-definite, and $\qA\succ 0$ means that $\qA$
is Hermitian positive definite.
$\mathrm{diag}(\qa)$ denotes a diagonal matrix with the elements of the vector $\qa$ along its diagonal.
$\|\qa\|$ denotes Euclidean norm of vector $\qa$.
$\qI_n$ denotes the identity matrix of order $n$ (the subscript is dropped when the dimension is obvious).
$\mathbb{E}\{\cdot\}$ denotes expectation operator.
In this paper, $\log(\cdot)$ denotes base-$e$ logarithm where $e= 2.71828\cdots$.

\begin{figure}[hbtp]
\centering
\includegraphics[width=3.3in]{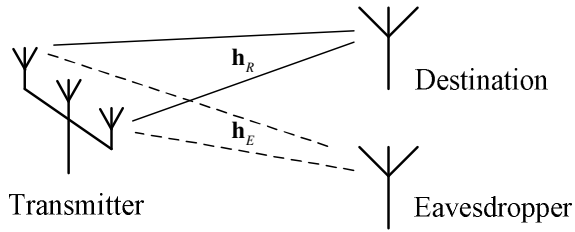}
\caption{System model.}
\label{systemmodel}
\end{figure}

\section{System Model and Problem Statement}\label{sec:2}

Consider a Gaussian MISO wiretap channel shown in Fig. \ref{systemmodel}, where the transmitter is equipped with $n_T$ antennas, while
the legitimate receiver and an eavesdropper each have a single antenna.
The received signals at the legitimate receiver and the eavesdropper are respectively given by
\al{
y_R&=\qh_R^\dagger\qx+v_R,\\
y_E&=\qh_E^\dagger\qx+v_E\label{system}
}
where $\qx$ is the $n_T\times 1$ transmitted signal vector with zero mean and $n_T \times n_T$ covariance matrix $\qR_x\succeq 0$,
i.e., $\qx\sim \mathcal{CN}(\mb{0}, \qR_x)$;
$\qh_R$, $\qh_E$ are respectively channel vectors between the transmitter and legitimate receiver,
and between the transmitter and eavesdropper;
$v_R\sim \mathcal{CN}(0,\sigma_v^2)$, $v_E\sim \mathcal{CN}(0,\sigma_v^2)$
are the noises at the legitimate receiver and the eavesdropper, respectively.
We can represent $\qR_x$ in terms of the average signal energy $E_s$ and
normalized signal covariance matrix $\qQ$, so that $\qR_x = E_s\qQ$
and $\mathrm{Tr}(\qQ) = 1$. The signal-to-noise ratio (SNR) is defined
as $\rho\triangleq E_s/\sigma_v^2$.

We assume that full CSI is available at
both the legitimate receiver and the eavesdropper, and
only the statistical information on  the eavesdropper channel is available at the transmitter.
We consider two cases, depending on the type of information available at the transmitter on the legitimate channel:
\begin{itemize}
  \item [a)] Only statistical information on the legitimate channel is available at the transmitter,
  i.e., the transmitter knows the distributions of $\qh_R$ and $\qh_E$ given by
$\qh_R\sim \mathcal{CN}(\mb{0}, {\bs \Sigma}_R)$, $\qh_E\sim \mathcal{CN}(\mb{0}, {\bs \Sigma}_E)$
with covariances ${\bs \Sigma}_R\succ 0$, and  ${\bs \Sigma}_E\succ 0$, respectively.
The ergodic secrecy capacity of the Gaussian MISO wiretap system (\ref{system}) equals \cite{Bhargava}
\equ{
C_s\triangleq\max_{\qQ\succeq 0, \mathrm{Tr}(\qQ)=1} C_s(\qQ)\label{ergoCap}
}
where $C_s(\qQ)$ is the ergodic secrecy rate given by
\al{
C_s(\qQ)&=\mathbb{E}_{\qh_R}\{\log (1+\rho\qh_R^\dagger\qQ\qh_R)\}\nnb\\
&\quad -\mathbb{E}_{\qh_E}\{\log (1+\rho\qh_E^\dagger\qQ\qh_E)\}.\label{ergoSecRate}
}
  \item [b)] Full CSI on the legitimate channel is available at the transmitter. The ergodic secrecy rate is given by \cite{Shafiee2}
\equ{
C_s(\qQ)=\log (1+\rho\qh_R^\dagger\qQ\qh_R)-\mathbb{E}_{\qh_E}\{\log (1+\rho\qh_E^\dagger\qQ\qh_E)\}.\label{ergoSecRateCSI}
}

\end{itemize}

The transmitter optimization problem is to find the optimal input covariance matrix $\qQ$
to maximize $C_s(\qQ)$ for cases a) and b).
We denote the feasible set as
$\Omega=\{\qQ|\qQ\succeq 0, \mathrm{Tr}(\qQ)=1\}$ which is a convex set.

The problem is of interest when a positive secrecy rate can be achieved, i.e., $C_s(\qQ)>0$ for some $\qQ$.
The conditions to ensure a positive ergodic capacity are provided in the following lemmas.

\medskip
\begin{lemma}\label{CondPosRate}
{\em For $\qh_R\sim \mathcal{CN}(\mb{0}, {\bs \Sigma}_R)$, the sufficient and necessary condition
under which $C_s(\qQ)>0$ for some $\qQ$ is that ${\bs \Sigma}_R - {\bs \Sigma}_E$ is non negative semi-definite.
}
\end{lemma}
The proof is given in Appendix \ref{ProofLemCondPosRate}.

\medskip
\begin{lemma}\label{CondPosRateCSI}
{\em When $\qh_R$ is completely known at the transmitter, a sufficient condition
under which $C_s(\qQ)>0$ for some $\qQ$ is that
$\qh_R\qh_R^\dagger-{\bs \Sigma}_E$ is non negative semi-definite.
}
\end{lemma}
The proof is given in Appendix \ref{ProofLemCondPosRateCSI}.

\section{Calculation of Ergodic Secrecy Rate}\label{sec:CsQ}

The calculation of the ergodic secrecy rate involves calculation of terms like $\mathbb{E}_{\qz}\{\log (1+\rho\qz^\dagger\qQ\qz)\}$
with $\qz \sim \mathcal{CN}(\mb{0}, \qR)$.
To this end, following the analysis of \cite[Eq. (64)]{Moustakas}, we give the following lemma.
The proof is given in Appendix \ref{ProofLem:ExpectCal}.
\medskip
\begin{lemma}\label{Lem:ExpectCal}
{\em Let $\qR^{1/2}\qQ\qR^{1/2}$ have a eigen-decomposition $\qU_1\qD_1\qU_1^\dagger$ where
$\qD_1=\mathrm{diag}(d_1,\cdots,d_M,0,\cdots,0)$, and
$d_1> \cdots > d_M > 0$ are the $M$ non-zero eigenvalues. For $\qz \sim \mathcal{CN}(\mb{0}, \qR)$, it holds:
\equ{
\mathbb{E}_{\qz}\{\log (1+\rho\qz^\dagger\qQ\qz)\}=\sum_{j=1}^M\frac{F_1(\rho d_j)}{\prod_{i\ne j}^M(1-d_i/d_j)} \label{E_z}
}
where $F_1(x)=e^{1/x}\mathrm{E}_1(1/x)$ with $\mathrm{E}_1(x)=\int_x^\infty \frac{e^{-t}}{t}\mathrm{d}t$ being the exponential integral.
}
\end{lemma}
\medskip

Based on (\ref{E_z}),  we can calculate $C_s(\qQ)$ by simply letting $\qR={\bs \Sigma}_R$ or $\qR={\bs \Sigma}_E$.

\section{Conditions for Optimal Input Covariance}\label{OptInpCov}

Next we obtain the necessary conditions for the optimal $\qQ$ by using the Karush-Kuhn-Tucker (KKT) conditions.
Let us construct the cost function
\equ{
L(\qQ, \theta, {\bs \Psi})=C_s(\qQ)-\theta (\mathrm{Tr}(\qQ)-1)+\mathrm{Tr}({\bs \Psi}\qQ)
}
where $\theta$ is the Lagrange multiplier associated with the constraint $\mathrm{Tr}(\qQ)=1$, and
${\bs \Psi}$ is the Lagrange multiplier associated with the constraint $\qQ\succeq 0$.
The KKT conditions enable us to write \cite{Boyd}
\al{
&{\bs \Theta}-\theta\qI_{n_T}+{\bs\Psi}=0,\label{KKTcond1}\\
&{\bs \Psi}\succeq 0, \mathrm{Tr}({\bs \Psi}\qQ)=0, \qQ\succeq 0, \mathrm{Tr}(\qQ)=1,\label{KKTcond2}
}
where ${\bs \Theta}=(\frac{\partial C_s(\qQ)}{\partial \qQ})^T$. By using the fact
$\frac{\partial \qh^\dagger\qQ\qh}{\partial \qQ}=(\qh\qh^\dagger)^T$, we have: for case a)
\equ{
{\bs \Theta}
=\mathbb{E}_{\qh_R}\bigg\{\frac{\rho\qh_R\qh_R^\dagger}{1+\rho\qh_R^\dagger\qQ\qh_R}\bigg\}
-\mathbb{E}_{\qh_E}\bigg\{\frac{\rho\qh_E\qh_E^\dagger}{1+\rho\qh_E^\dagger\qQ\qh_E}\bigg\}\label{Theta0}
}
and for case b)
\equ{
{\bs \Theta}
=\frac{\rho\qh_R\qh_R^\dagger}{1+\rho\qh_R^\dagger\qQ\qh_R}
-\mathbb{E}_{\qh_E}\bigg\{\frac{\rho\qh_E\qh_E^\dagger}{1+\rho\qh_E^\dagger\qQ\qh_E}\bigg\}.\label{Theta0CSI}
}

From the KKT conditions (\ref{KKTcond1}) and (\ref{KKTcond2}), we obtain the equivalent
(but without containing the Lagrange multipliers) conditions for optimal $\qQ$ consisting of a set of
equations given in the following theorem. Please see Appendix \ref{proofTheo1} for details.
\medskip
\begin{theorem}\label{theo:1}
{\em The optimal $\qQ\succeq 0$ satisfies
\al{
\qQ{\bs \Theta}&={\bs \Theta}\qQ=\mathrm{Tr}(\qQ{\bs \Theta})\qQ\label{basicEq1}\\
\lambda_{\max}({\bs\Theta})&=\mathrm{Tr}(\qQ{\bs \Theta}).\label{basicEq2}
}
}
\end{theorem}
\medskip

The above conditions imply  that for the  optimal $\qQ$,  $\qQ{\bs \Theta}$ is a scaled version  of $\qQ$.
Any $\qQ$ satisfying the conditions of Theorem \ref{theo:1} is called as KKT solution.
In \S\ref{Sec:Algo}, we propose an algorithm to search for the KKT solution.
${\bs \Theta}$ is an important variable for the transmitter optimization problem.
For the calculation of $\bs \Theta$, we give the following lemma. The proof is given in Appendix \ref{ProofLem:ThetaCalc}.

\medskip
\begin{lemma}\label{Lem:ThetaCalc}
{\em
For $\qz \sim \mathcal{CN}(\mb{0}, \qR)$, it holds
\equ{
\mathbb{E}_{\qz}\bigg\{\frac{\rho\qz\qz^\dagger}{1+\rho\qz^\dagger\qQ\qz}\bigg\}=\rho\qR^{1/2}\qU_1\qY\qU_1^\dagger\qR^{1/2}
\label{ExpectTheta}
}
where $\qY$ is a diagonal matrix with $(k,k)$th entry given by
\al{
Y_{kk}=& \sum_{j=1, j\ne k}^M\frac{F_1(\rho d_j)-F_1(\rho d_k)}
{(\rho d_j-\rho d_k)\prod_{i\ne j}^M(1-d_i/d_j)}\nnb\\
&\quad+\frac{F_2(\rho d_k)}{\prod_{i\ne k}^M(1-d_i/d_k)} , k\le M\\
Y_{kk}=&\sum_{j=1}^M\frac{F_1(\rho d_j)}{\rho d_j\prod_{i\ne j}^M(1-d_i/d_j)}
, k>M.
}
with $F_2(x)=\frac{1}{x}-\frac{1}{x^2}e^{1/x}\mathrm{E}_1(1/x)$, and $F_1(x)$, $\mathrm{E}_1(x)$, $\qU_1$, $d_i$, $M$ are defined
in Lemma \ref{Lem:ExpectCal}.
}
\end{lemma}
\medskip

Based on (\ref{ExpectTheta}), we can calculate ${\bs \Theta}$ by simply letting $\qR={\bs \Sigma}_R$ or $\qR={\bs \Sigma}_E$.

In the following, we show that for some special cases, more information about ${\bf Q}$ than that of Theorem 1 can be obtained.
%[LI: why is this useful if you have an algorithm to find Q based on Theorem 1???]
%[because the convergent solution (KKT solution) of the algorithm has not been shown to be the optimal

\subsection{ $\qh_R$ is completely known at the transmitter}\label{Sec:CaseB}

We will prove that if $C_s(\qQ)>0$ for some $\qQ$, then the optimal $\qQ$ always has rank one,
i.e., beamforming is optimal. We put the proof in the second part of the subsection.
In the first part of the subsection, we analyze how this result reduces our problem
to a problem of one variable.

Based on this result, we let $\qQ=\qu\qu^\dagger$ with $\|\qu\|^2=1$
and the problem is reduced to
\equ{
C_s(\qQ)=\log (1+\rho\qh_R^\dagger\qu\qu^\dagger\qh_R)-\mathbb{E}_{\qh_E}\{\log (1+\rho\qh_E^\dagger\qu\qu^\dagger\qh_E)\}\label{ergoSecRateCSIRankOne}
}
which, by using (\ref{ErgodicSecRate}) and (\ref{int1}), can be rewritten as
\equ{
C_s(\qQ)=\log (1+\rho\qu^\dagger\qh_R\qh_R^\dagger\qu) -F_1(\rho \qu^\dagger{\bs \Sigma}_E\qu).\label{ergoSecRateCSInT2}
}
Let $\qu^\dagger\qh_R\qh_R^\dagger\qu=z\|\qh_R\|^2$. Then $0\le z\le 1$.
Note that $F_1(x)$ is an increasing function. Thus, for fixed $z$, $\qu^\dagger{\bs \Sigma}_E\qu$
should be minimized. Define
\al{
\phi(z) = &\min_{\qu} \ \qu^\dagger{\bs \Sigma}_E\qu\label{phi_z}\\
          &\mathrm{s.t.}\quad \qu^\dagger\qh_R\qh_R^\dagger\qu=z\|\qh_R\|^2, \, \mathrm{and}\, \|\qu\|^2 =1.\nnb
}
Then, our problem is reduced to
\equ{
C_s(z)=\log (1+\rho \|\qh_R\|^2 z) -F_1(\rho \phi(z)), \ 0\le z\le 1.\label{ergoSecRateCSIReduced}
}
Since the problem of (\ref{phi_z}) belongs to the class of quadratically constrained quadratic programming (QCQP) with
two constraints, it can be exactly solved \cite{Huang}, and is equivalent to its semidefinite programming (SDP) relaxation, i.e.,
\al{
\phi(z) = &\min_{\qX} \ \mathrm{Tr}({\bs \Sigma}_E\qX)\label{phi_zSDP}\\
          &\mathrm{s.t.}\quad \mathrm{Tr}(\qh_R\qh_R^\dagger\qX)=z\|\qh_R\|^2, \, \mathrm{and}\, \mathrm{Tr}(\qX) =1,\nnb\\
          &\quad\quad\ \qX \succeq 0.\nnb
}
For any given $z$, the problem of (\ref{phi_zSDP}) is an SDP and can be effectively solved via CVX software \cite{Grant}.
\begin{lemma}\label{Lem:phi_zConvex}
{\em The function $\phi(z)$ is a convex function.
}
\end{lemma}
The proof is given in Appendix \ref{ProofLem:phi_zConvex}.

Since $\phi(z)$ is a convex function, according to well-known properties of convex functions,
we know that $\phi(z)$ is continuous and Lipschitz continuous \cite[Corollary 2.3.1]{Gromicho},
and is differentiable at all but at most countably many points (left and right derivatives always exists) \cite[Theorem 2.3.4]{Gromicho}.
Further study on $\phi(z)$ and proposing more effective method for
the optimization of $C_s(z)$ can be our future work.

For the special case ${\bs \Sigma}_E=\alpha\qI$ (${\bf h}_E$ has a trivial covariance),
(\ref{ergoSecRateCSInT2}) becomes
\equ{
C_s(\qQ)=\log (1+\rho\qu^\dagger\qh_R\qh_R^\dagger\qu) - F_1(\rho\alpha).\label{ergoSecRateIdCSI}
}
Obviously, the optimal $\qu=\qh_R/\|\qh_R\|$, the optimal $\qQ=\qh_R\qh_R^\dagger/\|\qh_R\|^2$ and
\equ{
[\max_{\qQ} C_s(\qQ)]=\log (1+\rho\|\qh_R\|^2) - F_1(\rho \alpha).\label{ergoSecRateIdCSI-b}
}
This case was considered in \cite{Shafiee2} and the above result is consistent with that in \cite{Shafiee2}.

\noindent{\em Remark}: If the optimal $\qu$ do not achieve $C_s(\qQ)>0$, then $C_s(\qQ)\le 0$ for any $\qQ$.
\medskip

In the remainder of the subsection, we give the proof for that
if $C_s(\qQ)>0$ for some $\qQ$, then the optimal $\qQ$ always has rank one.
We first provide a lemma that will be helpful in the following.
The proof is put in Appendix \ref{ProofLem:posiEig}.

\medskip
\begin{lemma}\label{Lem:posiEig}
{\em Let $\qA$ be a positive definite matrix,  $\qa$ be a vector. If $\qa\qa^\dagger-\qA$
has a positive eigenvalue, then it has all negative eigenvalues except for a positive eigenvalue.
}
\end{lemma}
\medskip

Via Lemma \ref{Lem:posiEig}, we can show that, if $C_s(\qQ)>0$, then ${\bs \Theta}$ has all negative eigenvalues
except for a positive eigenvalue.
To see why this is the case, recall that
\equ{
{\bs \Theta}
=\frac{\rho\qh_R\qh_R^\dagger}{1+\rho\qh_R^\dagger\qQ\qh_R}
-\mathbb{E}_{\qh_E}\bigg\{\frac{\rho\qh_E\qh_E^\dagger}{1+\rho\qh_E^\dagger\qQ\qh_E}\bigg\}.\label{Theta0CSInT2}
}
According to \S\ref{Subsec:DependenceFullCSI} (after Lemma \ref{Lem:Expecta}),
we know that, if $C_s(\qQ)>0$, then $\mathrm{Tr}({\bs \Theta}\qQ)>0$.
Thus, ${\bs \Theta}$ has at least a positive eigenvalue.
According to (\ref{ExpectTheta}), we know that the second term in the right hand side of (\ref{Theta0CSInT2}) is positive definite.
Note that the first term in the right hand side of (\ref{Theta0CSInT2}) has the form $\qa\qa^\dagger$.
Thus, the desired result follows directly from Lemma \ref{Lem:posiEig}.

From Theorem \ref{theo:1}, we know that the optimal $\qQ$ and its associated ${\bs \Theta}$ are commutable.
Thus, there exists a unitary matrix $\qU_0$ that simultaneously diagonalizes $\qQ$ and ${\bs \Theta}$. Let
${\bs \Lambda}_Q$ and ${\bs \Lambda}_{\bs \Theta}$ be the corresponding diagonal matrices.
From (\ref{basicEq1}) in Theorem \ref{theo:1}, we know that
\equ{
{\bs \Lambda}_Q{\bs \Lambda}_{\bs \Theta}=\mathrm{Tr}({\bs \Theta}\qQ){\bs \Lambda}_Q\label{basicEq1CSI}
}
or equivalently,
\equ{
({\bs \Lambda}_Q)_{kk}({\bs \Lambda}_{\bs \Theta})_{kk}=\mathrm{Tr}({\bs \Theta}\qQ)({\bs \Lambda}_Q)_{kk}, \ k=1,\cdots, n_T.
\label{basicEq1CSIa}
}
Since $\mathrm{Tr}({\bs \Theta}\qQ)>0$, it follows from (\ref{basicEq1CSIa}) that, if $({\bs \Lambda}_Q)_{kk}>0$,
then $({\bs \Lambda}_{\bs \Theta})_{kk}=\mathrm{Tr}({\bs \Theta}\qQ)>0$.
However,  ${\bs \Lambda}_{\bs \Theta}$ has all negative diagonal entries except for a positive one. Thus,
${\bs \Lambda}_Q$ has only one nonzero diagonal entry, i.e, the optimal $\qQ$ has rank one.

\subsection{Only statistical information on ${\bf h}_R$ available at the transmitter and ${\bs \Sigma}_E=\alpha\qI$}\label{sec:noCSI}

During this subsection, we assume that ${\bs \Sigma}_R$ has simple spectrum (all eigenvalues are distinct),
since multiple eigenvalues are rare for generic Hermitian matrices \cite[\S 4]{Tao}.
Let ${\bs \Sigma}_R$ have the eigen-decomposition ${\bs \Sigma}_R=\qV_R{\bs \Lambda}_R\qV_R^\dagger$
where ${\bs \Lambda}_R=\mathrm{diag}(\eta_1,\cdots,\eta_{n_T})$, $\eta_1>\eta_2>\cdots>\eta_{n_T}$.
We use Theorem \ref{theo:1} to show that the optimal $\qQ$ has the same eigenvectors as ${\bs \Sigma}_R$,
i.e., $\qV_R^\dagger\qQ\qV_R$ is diagonal, denoted by ${\bs \Lambda}=\mathrm{diag}(\zeta_1,\cdots,\zeta_{n_T})$.
Using this result, our problem is reduced to
\al{
C_s(\qQ)&=\mathbb{E}_{\qh_w}\{\log (1+\rho\qh_w^\dagger{\bs \Lambda}_R^{1/2}{\bs \Lambda}{\bs \Lambda}_R^{1/2}\qh_w)\}\nnb\\
&\quad -\mathbb{E}_{\qh_w}\{\log (1+\alpha\rho\qh_w^\dagger{\bs \Lambda}\qh_w)\}\nnb\\
&=\mathbb{E}_{\qh_w}\{\log (1+\rho\sum\nolimits_{i=1}^{n_T} \eta_i \zeta_i |h_{w,i}|^2)\}\nnb\\
&\quad-\mathbb{E}_{\qh_w}\{\log (1+\alpha\rho\sum\nolimits_{i=1}^{n_T} \zeta_i |h_{w,i}|^2)\}.\label{CsReI}
}
The power constraint is $\sum_{i=1}^{n_T} \zeta_i = 1$.
For the case $n_T=2$, (\ref{CsReI}) becomes
\al{
&C_s(\qQ)=\mathbb{E}_{\qh_w}\{\log (1+\rho \eta_1 \zeta_1 |h_{w,1}|^2+ \rho \eta_2 (1-\zeta_1)|h_{w,2}|^2)\}\nnb\\
& -\mathbb{E}_{\qh_w}\{\log (1+\alpha\rho\zeta_1 |h_{w,1}|^2 + \alpha\rho(1-\zeta_1)|h_{w,2}|^2)\}.\label{CsnT2}
}
The constraint is $0\le \zeta_1\le 1$. Similarly to \S\ref{sec:CsQ}, the expectations in (\ref{CsReI}) and (\ref{CsnT2})
can be expressed in explicit form.
\medskip

In the remainder of the subsection, we give the proof.
Let $\qQ$ have the eigen-decomposition $\qQ=\qV{\bs \Lambda}\qV^\dagger$ where ${\bs \Lambda}$ is diagonal.
According to Appendix \ref{ProofLem:ThetaCalc}, we express ${\bs \Theta}$ as
\equ{
{\bs \Theta}
=\mathbb{E}_{\qh_w}\bigg\{\frac{\rho{\bs \Sigma}_R^{\frac{1}{2}}\qh_w\qh_w^\dagger{\bs \Sigma}_R^{\frac{1}{2}}}{1+\rho\qh_w^\dagger
{\bs \Sigma}_R^{\frac{1}{2}}\qQ{\bs \Sigma}_R^{\frac{1}{2}}\qh_w}\bigg\}
-\mathbb{E}_{\qh_w}\bigg\{\frac{\alpha\rho\qh_w\qh_w^\dagger}{1+\alpha\rho\qh_w^\dagger\qQ\qh_w}\bigg\}.
}
Let ${\bs \Sigma}_R^{\frac{1}{2}}\qQ{\bs \Sigma}_R^{\frac{1}{2}}=\qA$ and let $\qA$ have
the eigen-decomposition $\qA=\qV_A{\bs \Lambda}_A\qV_A^\dagger$ where
\equ{
{\bs \Lambda}_A=\mathrm{diag}(\lambda_1\qJ_1, \lambda_2\qJ_2, \cdots, \lambda_K\qJ_K),\label{LambdaA}
}
$\lambda_1>\lambda_2>\cdots>\lambda_{K-1}>\lambda_K\ge 0$ are distinct eigenvalues, and $\qJ_k$'s are identity matrices.
By using the fact that $\qU\qh_w$ and $\qh_w$ have the identical distributions for any unitary matrix $\qU$,
we express ${\bs \Theta}$ as
\equ{
{\bs \Theta}
={\bs \Sigma}_R^{1/2}\qV_A\qY_0\qV_A^\dagger{\bs \Sigma}_R^{1/2}-\qV\qZ_0\qV^\dagger\label{Theta:1}
}
where
\al{
\qY_0&=\mathbb{E}_{\qh_w}\bigg\{\frac{\rho\qh_w\qh_w^\dagger}{1+\rho\qh_w^\dagger{\bs \Lambda}_A\qh_w}\bigg\},\label{Y0}\\
\qZ_0&=\mathbb{E}_{\qh_w}\bigg\{\frac{\alpha\rho\qh_w\qh_w^\dagger}{1+\alpha\rho\qh_w^\dagger{\bs \Lambda}\qh_w}\bigg\}.
}
Similarly to Appendix \ref{ProofLem:ThetaCalc}, it can be shown that $\qY_0$ and $\qZ_0$ are both diagonal.

Observe that $\qV\qZ_0\qV^\dagger$ and $\qQ$ are commutable.
With this, from Theorem \ref{theo:1}, we know that ${\bs \Theta}$ and $\qQ$ are commutable
which enables us to get
\equ{
{\bs \Sigma}_R^{1/2}\qV_A\qY_0\qV_A^\dagger{\bs \Sigma}_R^{1/2}\qQ
=\qQ{\bs \Sigma}_R^{1/2}\qV_A\qY_0\qV_A^\dagger{\bs \Sigma}_R^{1/2}.\label{ThetaQcommute}
}
By inserting ${\bs \Sigma}_R^{\frac{1}{2}}\qQ{\bs \Sigma}_R^{\frac{1}{2}}=\qV_A{\bs \Lambda}_A\qV_A^\dagger$ into
(\ref{ThetaQcommute}), we get
\equ{
{\bs \Sigma}_R\qV_A\qY_0{\bs \Lambda}_A\qV_A^\dagger
=\qV_A{\bs \Lambda}_A\qY_0\qV_A^\dagger{\bs \Sigma}_R.\label{ThetaQcommuteB}
}
Since ${\bs \Lambda}_A$ and $\qY_0$ are both  diagonal matrices, it holds that
$\qY_0{\bs \Lambda}_A={\bs \Lambda}_A\qY_0$. With this,
by inserting ${\bs \Sigma}_R=\qV_R{\bs \Lambda}_R\qV_R^\dagger$ into (\ref{ThetaQcommuteB}), we get
\equ{
{\bs \Lambda}_R\qV_0{\bs \Lambda}_A\qY_0\qV_0^\dagger
=\qV_0{\bs \Lambda}_A\qY_0\qV_0^\dagger{\bs \Lambda}_R\label{ThetaQcommute2}
}
where $\qV_0=\qV_R^\dagger\qV_A$. From (\ref{ThetaQcommute2}) and the assumption that all diagonal entries of
${\bs \Lambda}_R$ are distinct,
we know that $\qV_0{\bs \Lambda}_A\qY_0\qV_0^\dagger$ is a diagonal matrix \cite[Special matrices: diagonal]{Brookes}.
On the other hand, it follows from ${\bs \Sigma}_R^{\frac{1}{2}}\qQ{\bs \Sigma}_R^{\frac{1}{2}}=\qV_A{\bs \Lambda}_A\qV_A^\dagger$
that $\qQ={\bs \Sigma}_R^{-1/2}\qV_A{\bs \Lambda}_A\qV_A^\dagger{\bs \Sigma}_R^{-1/2}$ which, when combined with the fact that
${\bs \Sigma}_R=\qV_R{\bs \Lambda}_R\qV_R^\dagger$, results in
\equ{
\qQ=\qV_R{\bs \Lambda}_R^{-1/2}(\qV_0{\bs \Lambda}_A\qV_0^\dagger){\bs \Lambda}_R^{-1/2}\qV_R^\dagger.\label{Q}
}
Next, we show that $\qV_0{\bs \Lambda}_A\qV_0^\dagger$ is diagonal.
Since $\qV_0{\bs \Lambda}_A\qY_0\qV_0^\dagger$ is diagonal,  there exists a $\qV_1$,
which is the column-permuted version  of $\qV_0$, such that $\qV_1{\bs \Lambda}_A\qY_0\qV_1^\dagger={\bs \Lambda}_A\qY_0$,
or equivalently, $\qV_1{\bs \Lambda}_A\qY_0={\bs \Lambda}_A\qY_0\qV_1$.
We aim to prove that $\qV_1{\bs \Lambda}_A={\bs \Lambda}_A\qV_1$.
According to (\ref{LambdaA}), (\ref{Y0}), (\ref{IntYkk1}), (\ref{IntYkk2}),
it is not difficult to show that ${\bs \Lambda}_A\qY_0$ has the form of
\equ{
{\bs \Lambda}_A\qY_0=\mathrm{diag}(\lambda_1'\qJ_1, \lambda_2'\qJ_2, \cdots, \lambda_K'\qJ_K)\label{LambdaA1}
}
where $\lambda_1'>\lambda_2'>\cdots>\lambda_{K-1}'>0$, $\lambda_K'\ge 0$, and $\lambda_K'\ne \lambda_k'$ for $k=1,\cdots,K-1$.
From (\ref{LambdaA1}) and the fact that $\qV_1{\bs \Lambda}_A\qY_0={\bs \Lambda}_A\qY_0\qV_1$,
we know that $\qV_1$ has the form of $\qV_1 = \mathrm{diag}(\qA_1, \qA_2, \cdots, \qA_K)$, where each $\qA_k$ is the same size
as the corresponding $\qJ_k$ \cite[Special matrices: diagonal]{Brookes}.
Thus, it is easy to verify that $\qV_1{\bs \Lambda}_A={\bs \Lambda}_A\qV_1$, or equivalently,
$\qV_1{\bs \Lambda}_A\qV_1^\dagger={\bs \Lambda}_A$. Therefore, since $\qV_1$ is the column-permuted matrix of $\qV_0$,
it follows  that $\qV_0{\bs \Lambda}_A\qV_0^\dagger$ is diagonal.
With this, from (\ref{Q}), we know that the optimal $\qQ$ has the same eigenvectors as ${\bs \Sigma}_R$.

\section{Dependence of $C_s(\qQ)$ on $\rho$}\label{subSec:ImpactOnSNR}

In this section we investigate  how the SNR, $\rho$,  impacts  the ergodic secrecy rate.

\subsubsection{Full CSI on $\qh_R$ at the transmitter}\label{Subsec:DependenceFullCSI}

We first provide a lemma that will be helpful in the following. The proof is given in Appendix \ref{proofLem:Expecta}.

\medskip
\begin{lemma}\label{Lem:Expecta}
{\em For a positive constant $x$ and a positive random variable $Y$, the following fact holds:
\equ{
\log x > \mathbb{E}(\log Y) \Longrightarrow \frac{1}{x} < \mathbb{E}\left(\frac{1}{Y}\right).
}
Here, $\Longrightarrow$ means that the right side follows from the left side.
}
\end{lemma}
\medskip

By using Lemma \ref{Lem:Expecta}, we can prove that, if  $C_s(\qQ)>0$, then $\mathrm{Tr}({\bs \Theta}\qQ)>0$.
To see why this is the case, we let $x=1+\rho\qh_R^\dagger\qQ\qh_R$ and $Y=1+\rho\qh_E^\dagger\qQ\qh_E$ which enables us
to write
\al{
C_s(\qQ)&=\log x - \mathbb{E}(\log Y) \\
\mathrm{Tr}({\bs \Theta}\qQ)&=
\mathbb{E}\bigg(\frac{1}{Y}\bigg)-\frac{1}{x}.\label{Theta0CSInT2}}
The desired result follows from Lemma \ref{Lem:Expecta}.

Taking the derivative of $C_s(\qQ)$ with respect to $\rho$, we get:
\al{
\frac{\partial C_s(\qQ)}{\partial \rho} &=  \frac{\qh_R^\dagger\qQ\qh_R}{1+\rho\qh_R^\dagger\qQ\qh_R}-
\mathbb{E}_{\qh_E}\bigg\{\frac{\qh_E^\dagger\qQ\qh_E}{1+\rho\qh_E^\dagger\qQ\qh_E}\bigg\}\nnb\\
&=\frac{\mathrm{Tr}({\bs \Theta}\qQ)}{\rho}.
}
Based on the fact that, if $C_s(\qQ)>0$, then $\mathrm{Tr}({\bs \Theta}\qQ)>0$, we get that
if $C_s(\qQ)>0$, then $\frac{\partial C_s(\qQ)}{\partial \rho} > 0$.
Thus,
if $C_s(\qQ)>0$ for some $\qQ$, then more power should achieve larger secrecy rate.
In other words, we should use the maximum power.

\subsubsection{Statistical information on ${\bf h}_R$ at the transmitter}

In this case, we deal with the situation ${\bs \Sigma}_R \succeq {\bs \Sigma}_E$.
Taking the derivative of $C_s(\qQ)$ with respect to $\rho$, we get:
\al{
\frac{\partial C_s(\qQ)}{\partial \rho} &=
\frac{1}{\rho}\mathbb{E}_{\qh_w}\bigg\{\frac{1}{1+\rho\qh_w^\dagger{\bs \Sigma}_E^{1/2}\qQ{\bs \Sigma}_E^{1/2}\qh_w}\bigg\}\nnb\\
&\quad -\frac{1}{\rho}\mathbb{E}_{\qh_w}\bigg\{ \frac{1}{1+\rho\qh_w^\dagger{\bs \Sigma}_R^{1/2}\qQ{\bs \Sigma}_R^{1/2}\qh_w}\bigg\}.
}
Here, we have used (\ref{hR}) and (\ref{hE}) in Appendix \ref{ProofLemCondPosRate}.
According to Ostrowski theorem \cite[p. 224]{Horn}, we know that
if $\qA \succeq \qB$ and $\qB \succ 0$, then $\lambda_k(\qA^{1/2}\qQ\qA^{1/2})\ge \lambda_k(\qB^{1/2}\qQ\qB^{1/2})$,
where $\lambda_k(\cdot)$ denotes the $k$th eigenvalue arranged in decreasing order.
Since ${\bs \Sigma}_R \succeq {\bs \Sigma}_E$, similarly to the methodology in Appendix \ref{ProofLemCondPosRate},
it is easy to prove that $\frac{\partial C_s(\qQ)}{\partial \rho}>0$.
Thus, more power should achieve larger secrecy rate. In other words, we should use the maximum power.\\
{\em Remarks}: For the situation ${\bs \Sigma}_R \nsucceq {\bs \Sigma}_E$, whether or not
$C_s(\qQ)>0$ imply that $\frac{\partial C_s(\qQ)}{\partial \rho} > 0$ has not been proved. This
can be our future work.

\section{The Optimal ${\bf Q}$ Under High SNR} \label{Sec:HighSNR}

In this subsection, we give an analysis for high SNR, i.e., $\rho\to \infty$.
Our results show that for high SNR, the optimal $\qQ$ has rank one, i.e., beamforming is optimal.
The detailed analysis is given as follows.
%[LI: elaborate on this claim and provide a reference??]

\subsection{Full CSI about $\qh_R$ at the transmitter}

According to \S\ref{Sec:CaseB}, the optimal $\qQ$ always has rank one. Let $\qQ=\qu\qu^\dagger$ with $\|\qu\|^2=1$.

For high SNR, by using the fact that $\log(1+x)\approx \log x$ for large $x$, we write
\al{
C_s(\qQ) &\approx \log(\rho\qh_R^\dagger\qu\qu^\dagger\qh_R)-\mathbb{E}_{\qh_E}\{\log(\rho\qh_E^\dagger\qu\qu^\dagger\qh_E)\}\nnb\\
&= \log(\qu^\dagger\qh_R\qh_R^\dagger\qu) -
\mathbb{E}_{\qh_w}\{\log (\qh_w^\dagger{\bs \Sigma}_E^{1/2}\qu\qu^\dagger{\bs \Sigma}_E^{1/2}\qh_w)\}\nnb\\
&= \log(\qu^\dagger\qh_R\qh_R^\dagger\qu) - \mathbb{E}_{h_{w,1}}\{\log (\qu^\dagger{\bs \Sigma}_E\qu|h_{w,1}|^2)\}\nnb\\
&= \log(\qu^\dagger\qh_R\qh_R^\dagger\qu) - \log (\qu^\dagger{\bs \Sigma}_E\qu)-
\mathbb{E}\{\log |h_{w,1}|^2\}\nnb\\
&=\log \frac{\qu^\dagger\qh_R\qh_R^\dagger\qu}{\qu^\dagger{\bs \Sigma}_E\qu}+\gamma\nnb\\
&\le \log(\qh_R^\dagger{\bs \Sigma}_E^{-1}\qh_R)+\gamma\label{CSIhighSNR}
}
where we have used the fact that $\qU\qh_w$ and $\qh_w$ have  identical distributions for any unitary matrix $\qU$,
and $\mathbb{E}\{\log |h_{w,1}|^2\}=-\gamma$ where $\gamma=0.577216\cdots$ is the Euler's constant
(since $2|h_{w,1}|^2\sim \chi^2(2)$, i.e., the chi-square distribution with degree of freedom $2$).
In (\ref{CSIhighSNR}), the maximum is achieved when $\qu={\bs \Sigma}_E^{-1}\qh_R/\|{\bs \Sigma}_E^{-1/2}\qh_R\|$.

\subsection{Statistics information about ${\bf h}_R$ at the transmitter}

For high SNR, similarly, we write
\al{
&C_s(\qQ)\nnb\\
\approx\ & \mathbb{E}_{\qh_R}\{\log(\rho\qh_R^\dagger\qQ\qh_R)\}
-\mathbb{E}_{\qh_E}\{\log(\rho\qh_E^\dagger\qQ\qh_E)\}\nnb\\
=\ & \mathbb{E}_{\qh_w}\{\log (\qh_w^\dagger{\bs \Sigma}_R^{\frac{1}{2}}\qQ{\bs \Sigma}_R^{\frac{1}{2}}\qh_w)\}-
\mathbb{E}_{\qh_w}\{\log (\qh_w^\dagger{\bs \Sigma}_E^{\frac{1}{2}}\qQ{\bs \Sigma}_E^{\frac{1}{2}}\qh_w)\}\nnb\\
=\ &\mathbb{E}_{\qh_w}\{\log \sum\nolimits_{k=1}^{n_T} a_k |h_{w,k}|^2\}-
\mathbb{E}_{\qh_w}\{\log \sum\nolimits_{k=1}^{n_T} b_k |h_{w,k}|^2\}\nnb\\
=\ & \mathbb{E}_{\qh_w}\bigg\{\log \frac{\sum_{k=1}^{n_T} a_k |h_{w,k}|^2}{\sum_{k=1}^{n_T} b_k |h_{w,k}|^2}\bigg\}\label{statishighSNR}
}
where $a_k$'s and $b_k$'s are the eigenvalues of ${\bs \Sigma}_R^{\frac{1}{2}}\qQ{\bs \Sigma}_R^{\frac{1}{2}}$
and ${\bs \Sigma}_E^{\frac{1}{2}}\qQ{\bs \Sigma}_E^{\frac{1}{2}}$ arranged in decreasing order, respectively,
and we have used the fact that $\qU\qh_w$ and $\qh_w$ have the identical distributions for any unitary matrix $\qU$.
Let $p_1$ and $p_{n_T}$ be the largest and smallest eigenvalues of
${\bs \Sigma}_R^{\frac{1}{2}}{\bs \Sigma}_E^{-1}{\bs \Sigma}_R^{\frac{1}{2}}$.
By writing
\equ{
{\bs \Sigma}_R^{\frac{1}{2}}\qQ{\bs \Sigma}_R^{\frac{1}{2}}
={\bs \Sigma}_R^{\frac{1}{2}}{\bs \Sigma}_E^{-\frac{1}{2}}\left({\bs \Sigma}_E^{\frac{1}{2}}\qQ
{\bs \Sigma}_E^{\frac{1}{2}}\right){\bs \Sigma}_E^{-\frac{1}{2}}{\bs \Sigma}_R^{\frac{1}{2}}
}
and applying Ostrowski theorem \cite[p. 224]{Horn}, we have
\equ{
a_k =  b_k \theta_k, \ p_{n_T}\le \theta_k\le p_1 (k=1,\cdots,n_T).\label{eigenRelation}
}
Since $\mathrm{rank}({\bs \Sigma}_R^{\frac{1}{2}}\qQ{\bs \Sigma}_R^{\frac{1}{2}})
=\mathrm{rank}({\bs \Sigma}_E^{\frac{1}{2}}\qQ{\bs \Sigma}_E^{\frac{1}{2}})$, the number of non-zero elements
of $a_k$' and $b_k$'s is the same,  denoted by $n$. Thus, from (\ref{eigenRelation}), we have
\equ{
\left[\max_{k=1,\cdots,n}\ \frac{a_k}{b_k}\right] \le p_1 =
\lambda_{\max}({\bs \Sigma}_R^{\frac{1}{2}}{\bs \Sigma}_E^{-1}{\bs \Sigma}_R^{\frac{1}{2}}).\label{Inequ}
}
To proceed, we need the following lemma.
\begin{lemma}\label{Lem:Inequ}
{\em For $x_k>0$, $y_k>0$, $k=1,\cdots,n$, it holds
\equ{
\frac{x_1+\cdots+x_n}{y_1+\cdots+y_n}\le \max_{k=1,\cdots,n}\ \frac{x_k}{y_k}.
}
}
\end{lemma}
The proof is simple: it is based on the following
\equ{
\frac{a+b}{c+d} \le \max\big\{\frac{a}{c}, \frac{b}{d}\big\},\ \forall a, b, c, d > 0.
}
From (\ref{Inequ}), using Lemma \ref{Lem:Inequ}, we get that: for any $\qh_w\ne 0$,
\al{
\frac{\sum_{k=1}^{n_T} a_k |h_{w,k}|^2}{\sum_{k=1}^{n_T} b_k |h_{w,k}|^2} &=
\frac{\sum_{k=1}^n a_k |h_{w,k}|^2}{\sum_{k=1}^n b_k |h_{w,k}|^2}\nnb\\
&\le \max_{k=1,\cdots,n}\ \frac{a_k}{b_k}\nnb\\
&\le \lambda_{\max}({\bs \Sigma}_R^{\frac{1}{2}}{\bs \Sigma}_E^{-1}{\bs \Sigma}_R^{\frac{1}{2}}).\label{RatioMax}
}
In (\ref{RatioMax}), the maximum is achieved {\em simultaneously} for any $\qh_w\ne 0$
when $\qQ=\qu_0\qu_0^\dagger$, $\qu_0={\bs \Sigma}_E^{-\frac{1}{2}}\qx_0/\|{\bs \Sigma}_E^{-\frac{1}{2}}\qx_0\|$
with $\qx_0$ being the eigenvector associated
with the largest eigenvalue of ${\bs \Sigma}_E^{-\frac{1}{2}}{\bs \Sigma}_R{\bs \Sigma}_E^{-\frac{1}{2}}$, and
correspondingly, $a_1=\qu_0^\dagger{\bs \Sigma}_R\qu_0$, $b_1=\qu_0^\dagger{\bs \Sigma}_E\qu_0$,
$a_2=\cdots=a_{n_T}=0$, $b_2=\cdots=b_{n_T}=0$, i.e., $n=1$.
To see why this is the case, noting that
$\lambda_{\max}({\bs \Sigma}_R^{\frac{1}{2}}{\bs \Sigma}_E^{-1}{\bs \Sigma}_R^{\frac{1}{2}})
=\lambda_{\max}({\bs \Sigma}_E^{-\frac{1}{2}}{\bs \Sigma}_R{\bs \Sigma}_E^{-\frac{1}{2}})$,
it is easy to verify that
\equ{
\frac{a_1}{b_1}=
\frac{\qx_0^\dagger{\bs \Sigma}_E^{-\frac{1}{2}}{\bs \Sigma}_R{\bs \Sigma}_E^{-\frac{1}{2}}\qx_0}{\qx_0^\dagger\qx_0}
=\lambda_{\max}({\bs \Sigma}_E^{-\frac{1}{2}}{\bs \Sigma}_R{\bs \Sigma}_E^{-\frac{1}{2}}).
}
The desired result follows.
Now, combining (\ref{statishighSNR}) and (\ref{RatioMax}), we have
\al{
C_s(\qQ) &\approx
\mathbb{E}_{\qh_w}\bigg\{\log \frac{\sum_{k=1}^{n_T} a_k |h_{w,k}|^2}{\sum_{k=1}^{n_T} b_k |h_{w,k}|^2}\bigg\}\nnb\\
&\le \log (\lambda_{\max}({\bs \Sigma}_R^{\frac{1}{2}}{\bs \Sigma}_E^{-1}{\bs \Sigma}_R^{\frac{1}{2}})).\label{statisHighSNRopt}
}
In (\ref{statisHighSNRopt}), the maximum is achieved when $\qQ=\qu_0\qu_0^\dagger$. Thus, the optimal $\qQ$ has rank one.

\section{Fixed Point Iteration}\label{Sec:Algo}

In this section we propose an algorithm to search for the KKT solution according to Theorem \ref{theo:1}.
When ${\bs \Theta}$ and $\qQ$ commute,
${\bs \Theta}+\gamma \qI_{n_T}$ and $\qQ$ commute for any real number $\gamma$, and vice versa.
Let $\gamma=(1+\beta)\max\{0, -\lambda_{\min}(\bs\Theta)\}$, $\beta>0$
and let $\qK={\bs \Theta}+\gamma\qI_{n_T}$. It holds that  $\qK\succ 0$.
From (\ref{basicEq1}), we get
\equ{
\qK\qQ=\mathrm{Tr}(\qK\qQ)\qQ.\label{K}
}
Equation (\ref{K}) looks like the eigenvalue equation $\qA\qx=\lambda\qx$, where
$\mathrm{Tr}(\qK\qQ)$ is the eigenvalue and $\qQ$ is the  corresponding eigenvector.
Recall that the power iteration method is a classical method for computing the
eigenvector associated with the largest eigenvalue of a matrix \cite[p. 533]{Meyer}
\equ{
\qx_{k+1}=\frac{\qA\qx_k}{\|\qA\qx_k\|}, k=0, 1, \cdots.\label{PowerIteration}
}
We can derive the similar algorithm. Note that there is a difference between (\ref{K}) and the eigenvalue
equation: $\qQ$ is a Hermitian matrix. Thus, the iteration (\ref{PowerIteration}) cannot be used directly.
From (\ref{K}), since $\qK$ and $\qQ$ commute, thus, we have that $\qK\qQ=\qK^{1/2}\qQ\qK^{1/2}$ and hence
\equ{
\qQ=\frac{\qK^{1/2}\qQ\qK^{1/2}}{\mathrm{Tr}(\qK^{1/2}\qQ\qK^{1/2})}
\triangleq f(\qQ).\label{mapping}
}
Note that $f(\qQ)\succeq 0$ and $\mathrm{Tr}(f(\qQ))=1$ for any $\qQ\in\Omega$.
The equation (\ref{mapping}) defines a mapping from a convex set to itself: $\Omega \to \Omega$, $\qQ \mapsto f(\qQ)$.
The optimal $\qQ$ corresponds to a fixed point of $f(\qQ)$, i.e., $f(\qQ^\circ)=\qQ^\circ$.
To search for the KKT solution, the iterative expression is
\equ{
\qQ^{k+1}=f(\qQ^{k}), k=0,1,\cdots
}
The initial point $\qQ^0$ can be set to $\qI_{n_T}/n_T$, or any $\qQ^0\in \Omega$.
The iterations stop when the relative error of $C_s(\qQ)$ in the successive iterations is less than a preset value,
e.g., $10^{-3}$ or $10^{-6}$.
If the convergent $\qQ$ satisfies (\ref{basicEq2}),
we obtain a KKT solution,
otherwise, we choose a different initial point.

\section{Numerical Simulations}\label{sec:sim}

In this section we provide some examples to illustrate the theoretical findings.
We assume that ${\bs \Sigma}_R$ is normalized as $\mathrm{Tr}({\bs \Sigma}_R)=n_T$, and
${\bs \Sigma}_E$ is multiplied correspondingly by a factor $\eta$.
In simulations, we assume that the correlation matrices of legitimate and eavesdropper channels follow
the Jakes' correlation model \cite{Jakes}, i.e., for $p, q = 1,\cdots, n_T$
\al{
{\bs \Sigma}_R(p, q)&=\mathcal{J}_0\big(\phi_R |p-q| 2\pi d/\lambda\big),\nnb\\
{\bs \Sigma}_E(p, q)&=\eta\mathcal{J}_0\big(\phi_E |p-q| 2\pi d/\lambda\big)\nnb
}
where $\mathcal{J}_0(\cdot)$
is the zero-order Bessel function of the first kind, $d$ is the element spacing,
$\lambda$ is the wavelength, and $\phi_R$ (or $\phi_E$) is a parameter that controls the
correlation among antennas and has its value determined by the distance between
the transmitter and receiver, and the incident angle of the
wavefront.
We set $d/\lambda=1/2$.

\subsection{The transmitter has full information about the legitimate channel and only statistical information about the eavesdropper channel}

We consider a MISO wiretap channel where $n_T=4$, $n_R=n_E=1$.
We set $\qh_R=[0.4282 + 0.0403\mathrm{i}, 0.8956 + 0.6771\mathrm{i}, 0.7310 + 0.5689\mathrm{i}, 0.5779 - 0.2556\mathrm{i}]^T$
and $\phi_E = 0.3$, $\eta=0.3$.
The eigenvalues of $\qh_R\qh_R^\dagger-{\bs\Sigma}_E$ are $2.0946, -0.0020, -0.1584, -0.4315$.

\begin{figure}[hbtp]
\centering
\includegraphics[width=4in]{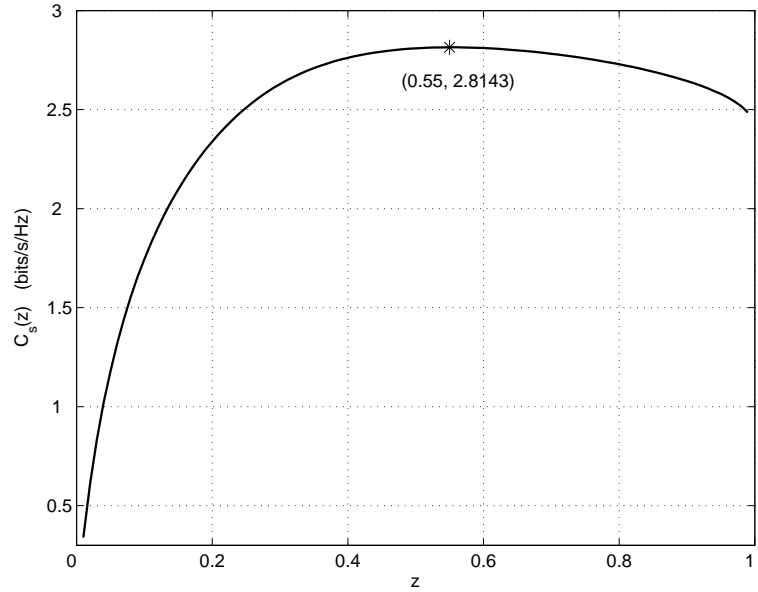}
\caption{The function $C_s(z)$, $z=0.01:0.01:0.99$. Full CSI on ${\bf h}_R$ is used,
$\mathrm{SNR}=10\,\mathrm{dB}$.}
\label{fig:Cs_z}
\end{figure}

Fig. \ref{fig:Cs_z} depicts the function $C_s(z)$ defined in (\ref{ergoSecRateCSIReduced})
for $z$ in $[0.01,0.99]$ with step $0.01$
and $\mathrm{SNR}=10\,\mathrm{dB}$.
Among these points, the optimal point is $(0.55, 2.8413)$ also depicted in Fig. \ref{fig:Cs_z}.

\begin{figure}[hbtp]
\centering
\includegraphics[width=4in]{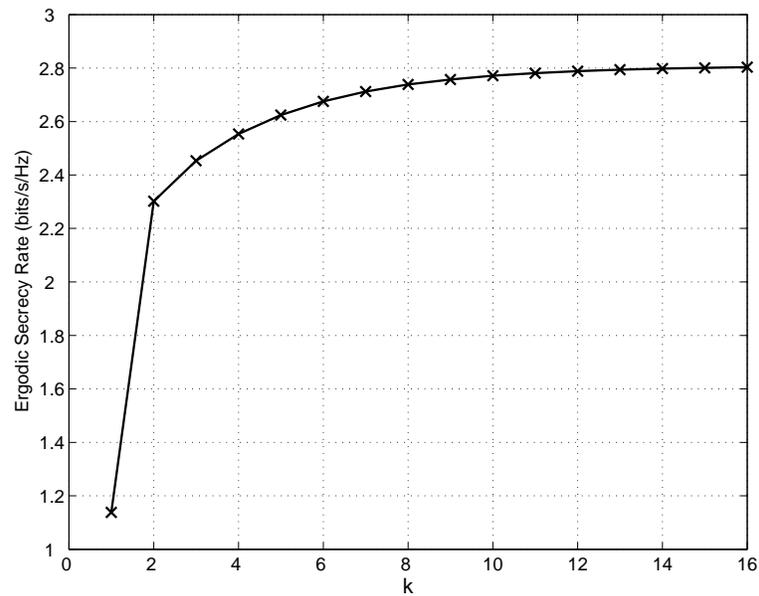}
\caption{Ergodic secrecy rate during the iteration. Full CSI on ${\bf h}_R$ is used; $\qQ^0=\frac{1}{4}\qI_4$,
$\mathrm{SNR}=10\,\mathrm{dB}$.}
\label{fig:6}
\end{figure}

Fig. \ref{fig:6} depicts the ergodic secrecy rates during the iteration
of the algorithm of Section \S\ref{Sec:Algo}
for $\mathrm{SNR}=10\,\mathrm{dB}$ and $\qQ^0=\frac{1}{4}\qI_4$.
The convergent ergodic secrecy rate is $2.8413$.
We can see that the algorithm converges rapidly. If we do $300$ iterations
for $\mathrm{SNR}=10\,\mathrm{dB}$ and $\qQ^0=\frac{1}{4}\qI_4$, the convergent values are:
\al{
\lambda(\qQ)&=\{1.0000, 0.0000, 0.0000, 0.0000\},\nnb\\
\lambda({\bs \Theta})&=\{0.4385, -0.0105, -1.3155, -2.2006\}\nnb\\
\mathrm{Tr}({\bs \Theta}\qQ)&=0.4385.\nnb
}

\begin{figure}[hbtp]
\centering
\includegraphics[width=4in]{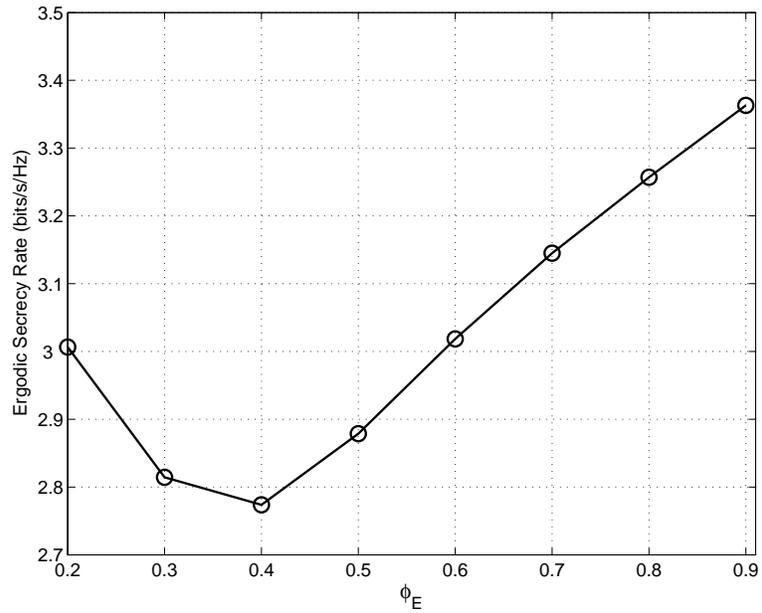}
\caption{Ergodic secrecy rate for different values of $\phi_E$. Full CSI on ${\bf h}_R$ is used; $\qQ^0=\frac{1}{4}\qI_4$,
$\mathrm{SNR}=10\,\mathrm{dB}$.}
\label{fig:8}
\end{figure}

\begin{figure}[hbtp]
\centering
\includegraphics[width=4in]{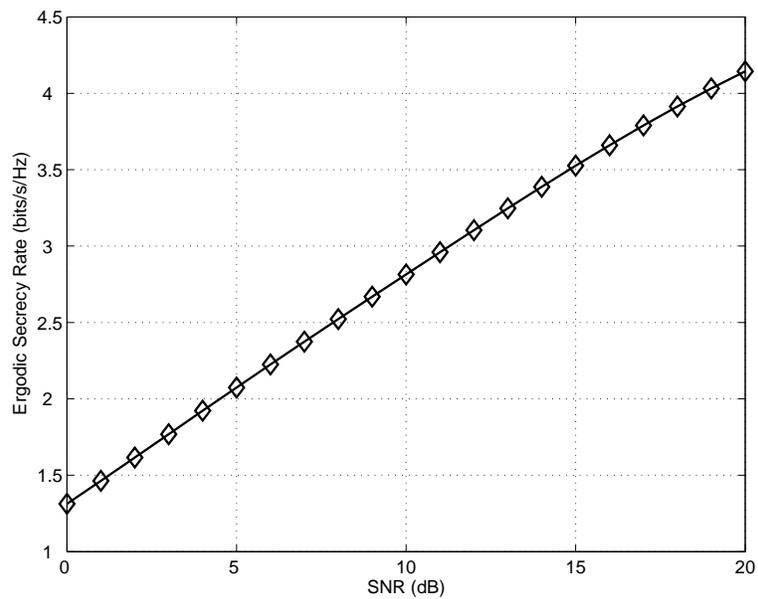}
\caption{Ergodic secrecy rate for different $\mathrm{SNR}$. Full CSI on ${\bf h}_R$ is used.}
\label{fig:9}
\end{figure}

Fig. \ref{fig:8} plots the ergodic secrecy rates for different $\phi_E$ from $0.2$ to $0.9$.
It can be seen from Fig. \ref{fig:8} that the ergodic secrecy rate decreases first, and then increases with $\phi_E$.
Fig. \ref{fig:9} plots the ergodic secrecy rates for different $\mathrm{SNR}$ and $\phi_E = 0.3$.
As is revealed in \S\ref{subSec:ImpactOnSNR}, the ergodic secrecy
rate increases with $\mathrm{SNR}$.

\subsection{The transmitter has  only statistical information about both the legitimate channel and the eavesdropper channel}

We set $\phi_R = 0.5$, $\phi_E = 0.3$, $\eta=0.3$.
The eigenvalues of ${\bs\Sigma}_R-{\bs\Sigma}_E$ are $1.3503, 0.9848, 0.4432, 0.0217$.

\begin{figure}[hbtp]
\centering
\includegraphics[width=4in]{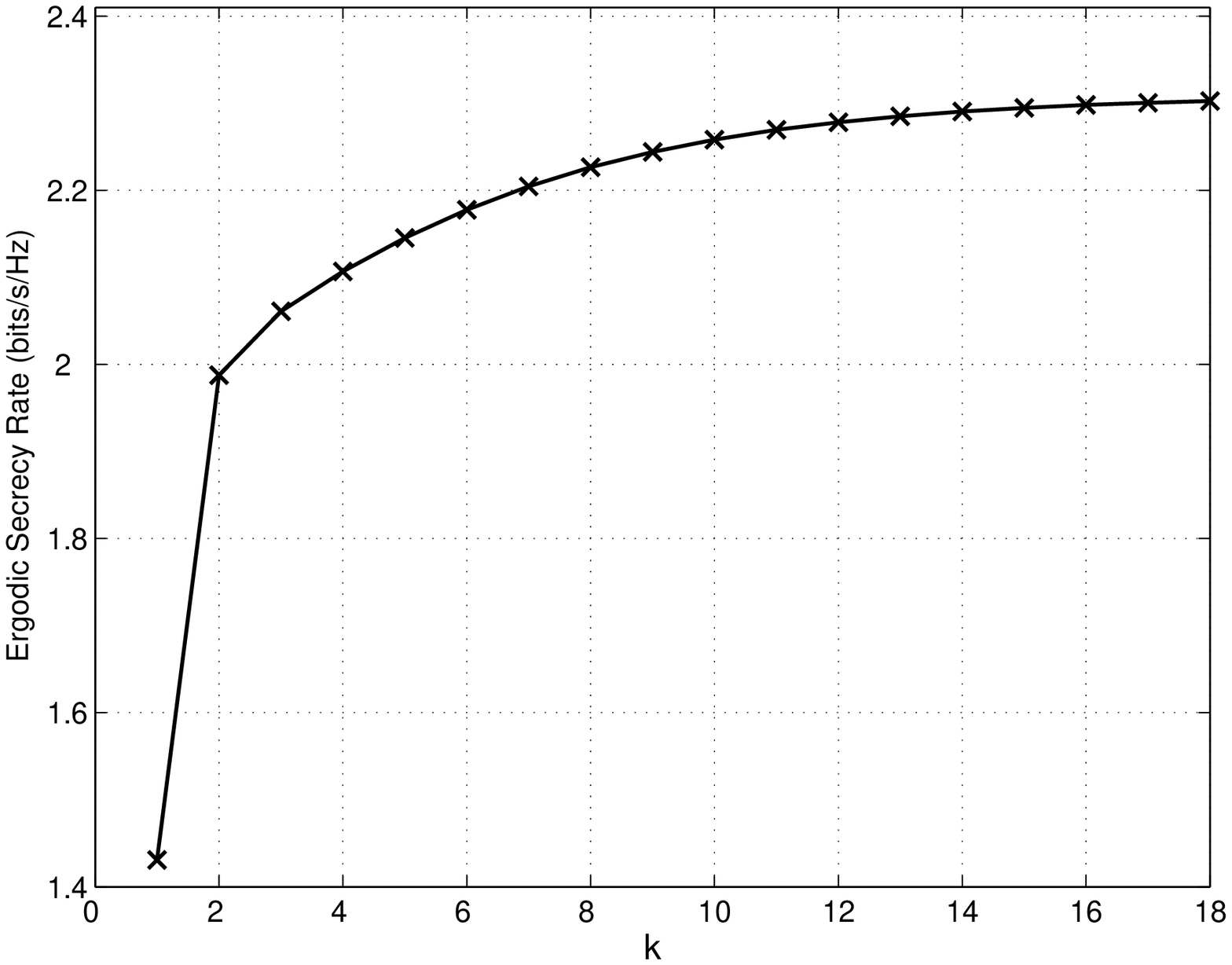}
\caption{Ergodic secrecy rate during the iteration. Only statistical info. on ${\bf h}_R$ is used;
$\qQ^0=\frac{1}{4}\qI_4$, $\mathrm{SNR}=10\,\mathrm{dB}$.}
\label{fig:1}
\end{figure}

\begin{figure}[hbtp]
\centering
\includegraphics[width=4in]{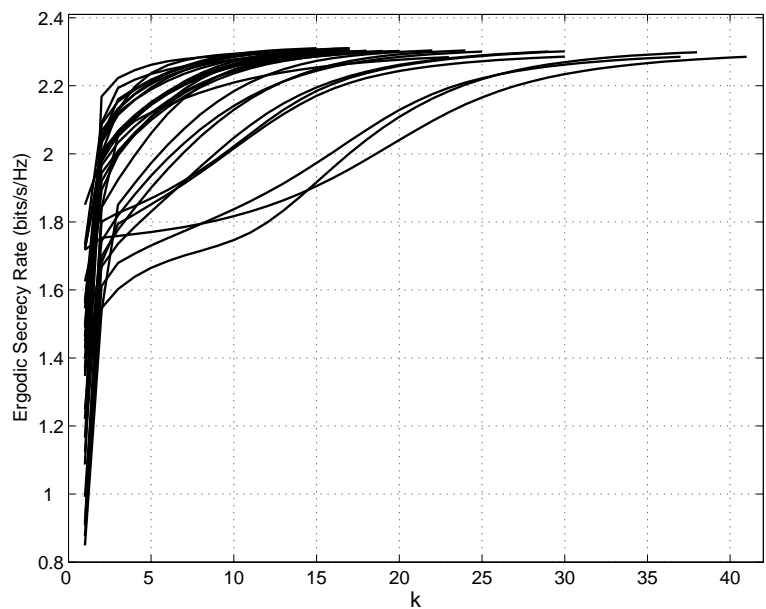}
\caption{Ergodic secrecy rate during the iteration. Only statistical info. on ${\bf h}_R$ is used;
$30$ random $\qQ^0$, $\mathrm{SNR}=10\,\mathrm{dB}$.}
\label{fig:2}
\end{figure}

Fig. \ref{fig:1} depicts the ergodic secrecy rates during the iteration for $\mathrm{SNR}=10\,\mathrm{dB}$ and $\qQ^0=\frac{1}{4}\qI_4$,
while Fig. \ref{fig:2} depicts the ergodic secrecy rates for $30$ random $\qQ^0\in \Omega$.
We can see that the algorithm converges rapidly. If we do $300$ iterations
for $\mathrm{SNR}=10\,\mathrm{dB}$ and $\qQ^0=\frac{1}{4}\qI_4$, the convergent values are:
\al{
\lambda(\qQ)&=\{0.5129, 0.4871, 0.0000, 0.0000\},\nnb\\
\lambda({\bs \Theta})&=\{0.7452, 0.7452, -2.5972, -5.4438\}\nnb\\
\mathrm{Tr}({\bs \Theta}\qQ)&=0.7452.\nnb
}
We can see that the convergent $\qQ$ has rank two.

\begin{figure}[hbtp]
\centering
\includegraphics[width=4in]{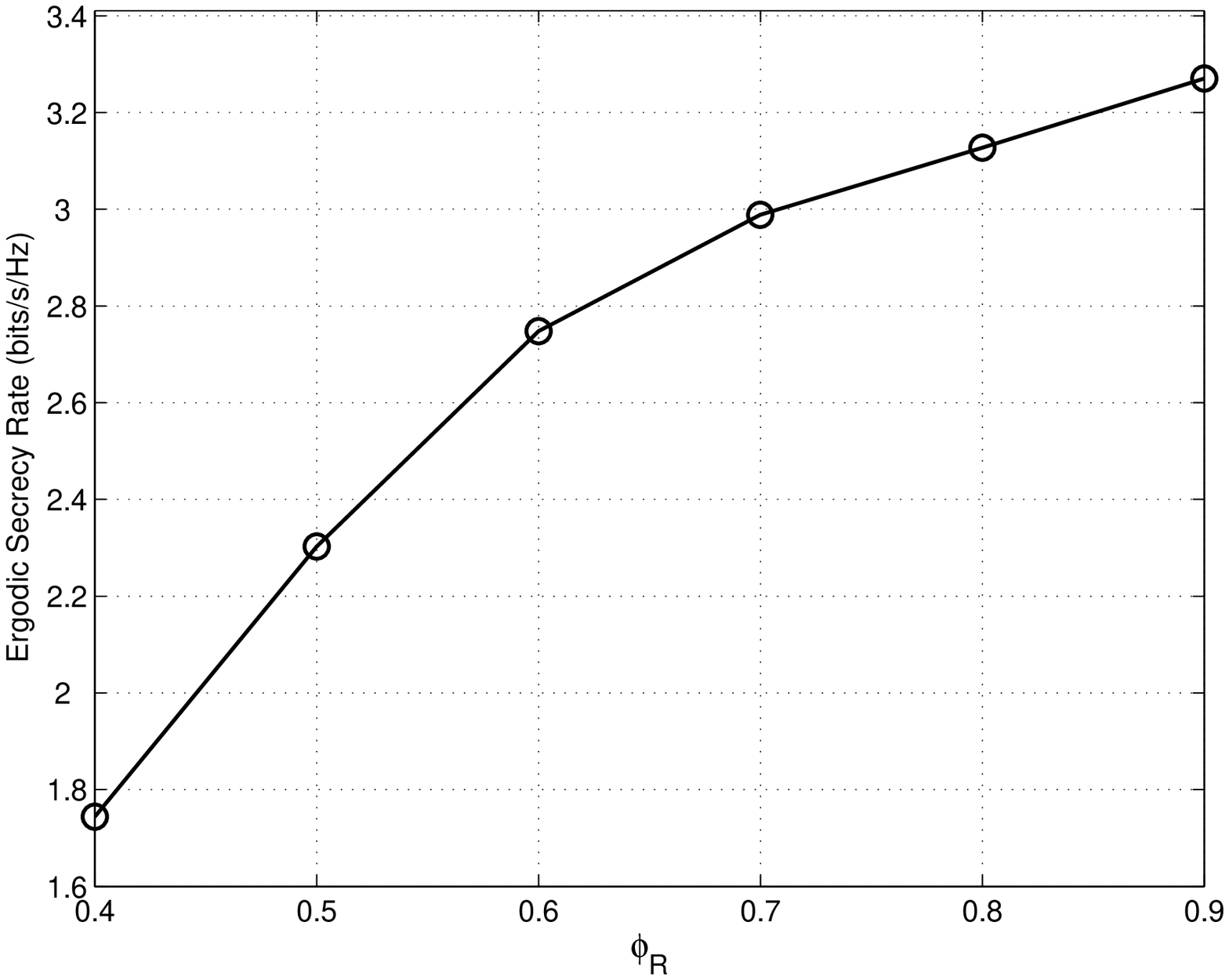}
\caption{Ergodic secrecy rate for different values of $\phi_R$. Only statistical info. on ${\bf h}_R$ is used;
$\qQ^0=\frac{1}{4}\qI_4$, $\mathrm{SNR}=10\,\mathrm{dB}$.}
\label{fig:3}
\end{figure}

\begin{figure}[hbtp]
\centering
\includegraphics[width=4in]{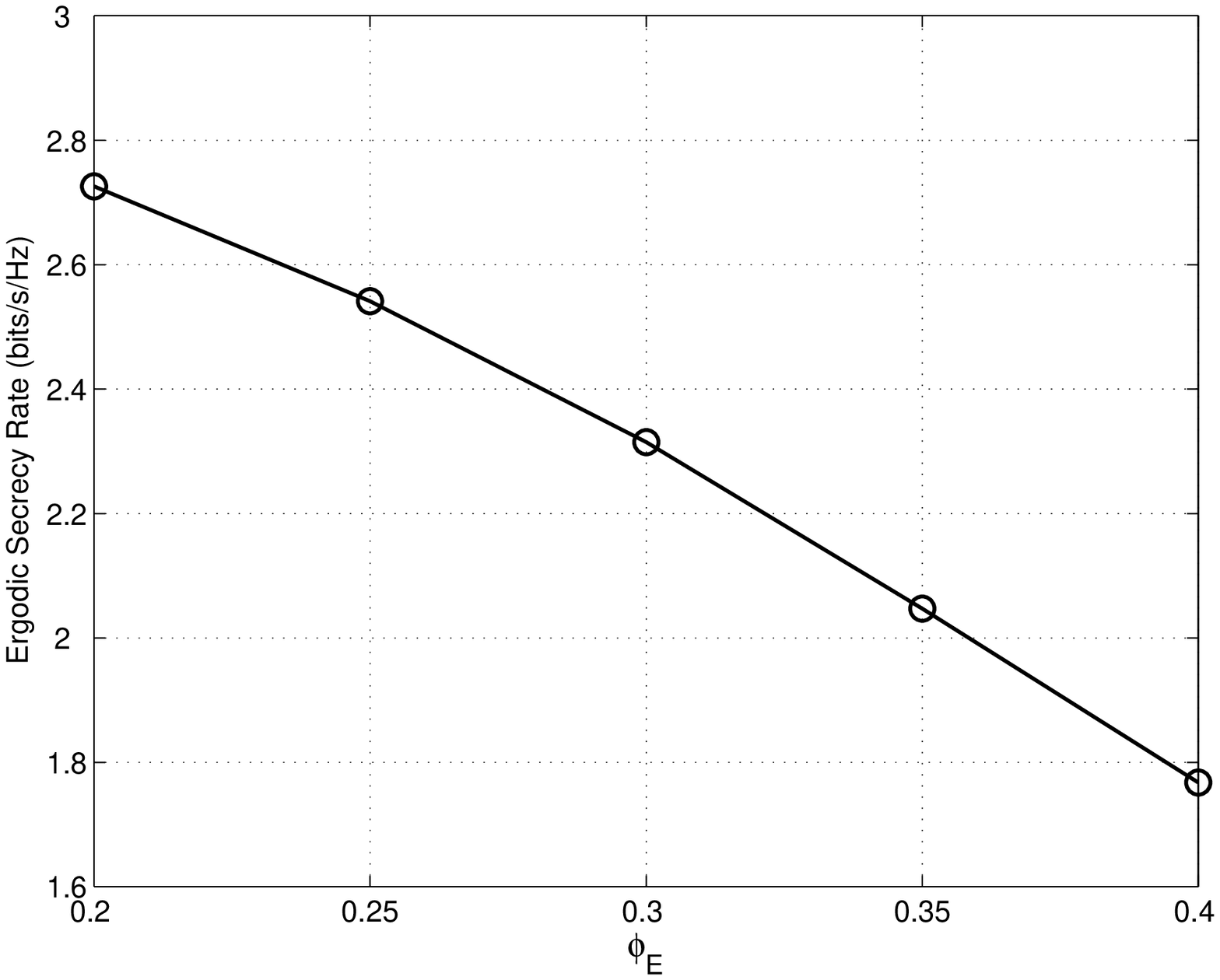}
\caption{Ergodic secrecy rate for different values of $\phi_E$. Only statistical info. on ${\bf h}_R$ is used; $\qQ^0=\frac{1}{4}\qI_4$,
$\mathrm{SNR}=10\,\mathrm{dB}$.}
\label{fig:4}
\end{figure}

\begin{figure}[hbtp]
\centering
\includegraphics[width=4in]{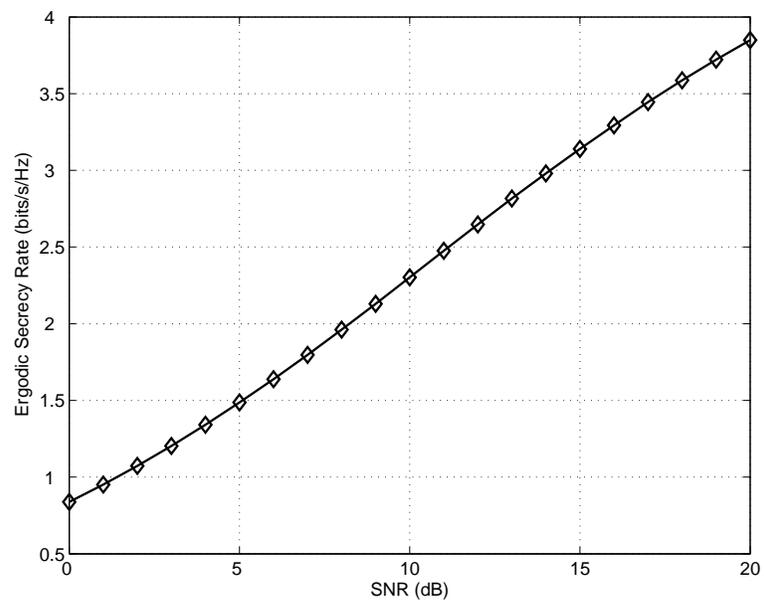}
\caption{Ergodic secrecy rate for different $\mathrm{SNR}$. Only statistical info. on ${\bf h}_R$ is used.}
\label{fig:5}
\end{figure}

Fig. \ref{fig:3} plots the ergodic secrecy rates for different $\phi_R$ from $0.4$ to $0.9$.
It can be seen from Fig. \ref{fig:3} that the ergodic secrecy rate increases with $\phi_R$.
Fig. \ref{fig:4} plots the ergodic secrecy rates for different $\phi_E$ from $0.2$ to $0.4$.
Fig. \ref{fig:5} plots the ergodic secrecy rates for different $\mathrm{SNR}$.
As is revealed in \S\ref{subSec:ImpactOnSNR}, when ${\bs\Sigma}_R\succ{\bs\Sigma}_E$, the ergodic secrecy
rate increases with $\mathrm{SNR}$.

\section{conclusion}\label{sec:conclusion}

We have investigated the problem of finding the optimal input covariance matrix that achieves ergodic secrecy capacity subject to a power constraint. We extend the existing result to nontrivial covariances of the legitimate and eavesdropper channels.
We have derived the necessary conditions for the optimal input covariance matrix in the form of a set of equations and
propose an algorithm to solve the equations.

\appendices

\section{Proof of Lemma \ref{CondPosRate}}\label{ProofLemCondPosRate}

We prove the result in two parts. \underline{First}, we prove that if ${\bs \Sigma}_E - {\bs \Sigma}_R \succeq 0$,
then $C_s \le 0$.
Since $\qh_R\sim \mathcal{CN}(\mb{0}, {\bs \Sigma}_R)$, $\qh_E\sim \mathcal{CN}(\mb{0}, {\bs \Sigma}_E)$,
we can write
\al{
\qh_R&={\bs \Sigma}_R^{1/2}\qh_w, \label{hR}\\
\qh_E&={\bs \Sigma}_E^{1/2}\qh_w\label{hE}.
}
By inserting (\ref{hR}) and (\ref{hE}) into (\ref{ergoSecRate}), we get
\al{
C_s(\qQ)&=\mathbb{E}_{\qh_w}\{\log (1+\rho\qh_w^\dagger{\bs \Sigma}_R^{1/2}\qQ{\bs \Sigma}_R^{1/2}\qh_w)\}\nnb\\
&\quad -\mathbb{E}_{\qh_w}\{\log (1+\rho\qh_w^\dagger{\bs \Sigma}_E^{1/2}\qQ{\bs \Sigma}_E^{1/2}\qh_w)\}.
}
Let $x_1\ge x_2 \ge \cdots \ge x_{n_T}$ and $y_1\ge y_2\ge\cdots\ge y_{n_T}$ be
eigenvalues of ${\bs \Sigma}_R^{1/2}\qQ{\bs \Sigma}_R^{1/2}$ and ${\bs \Sigma}_E^{1/2}\qQ{\bs \Sigma}_E^{1/2}$,
respectively. By using the fact that $\qU\qh_w$ and $\qh_w$ have the identical distributions for any unitary matrix $\qU$,
we have
\al{
C_s(\qQ)&=\mathbb{E}_{\qh_w}\bigg\{\log \bigg(1+\rho\sum_{i=1}^{n_T} x_i|h_{w, i}|^2\bigg)\bigg\}\nnb\\
&\quad -\mathbb{E}_{\qh_w}\bigg\{\log \bigg(1+\rho\sum_{i=1}^{n_T} y_i|h_{w, i}|^2\bigg)\bigg\}.
}
According to Ostrowski theorem \cite[p. 224]{Horn}, we know that
if $\qA \succeq \qB$ and $\qB \succ 0$, then $\lambda_k(\qA^{1/2}\qQ\qA^{1/2})\ge \lambda_k(\qB^{1/2}\qQ\qB^{1/2})$, where $\lambda_k(\cdot)$ denotes the $k$th eigenvalue arranged in decreasing order. With this, we know that
$x_i\le y_i$, $i=1,\cdots,n_T$. On the other hand, it is easy to verify that the following function
\equ{
g(z_1,\cdots,z_{n_T})=\mathbb{E}_{\qh_w}\bigg\{\log \bigg(1+\rho\sum_{i=1}^{n_T} z_i|h_{w, i}|^2\bigg)\bigg\}\label{func_g}
}
is strictly increasing with respect to $z_i$, $i=1,\cdots, n_T$. Thus, we get that $C_s(\qQ)\le 0$ for any $\qQ$.
This completes the first part.

\underline{Second}, we prove that if ${\bs \Sigma}_R - {\bs \Sigma}_E$ is none negative semi-definite, then
there exists a $\qQ$ such that $C_s(\qQ)>0$. Let $\qu$ be the eigenvector associated with the largest eigenvalue $\lambda$ of ${\bs \Sigma}_R - {\bs \Sigma}_E$. Since $\lambda>0$, we get that $\qu^\dagger({\bs \Sigma}_R - {\bs \Sigma}_E)\qu=\lambda>0$
and $\qu^\dagger{\bs \Sigma}_R\qu > \qu^\dagger{\bs \Sigma}_E\qu$.
We will prove that $\qQ=\qu\qu^\dagger$ achieves $C_s(\qQ)>0$.
In this case, we know that $x_1=\qu^\dagger{\bs \Sigma}_R\qu$, $x_2=\cdots=x_{n_T}=0$, $y_1=\qu^\dagger{\bs \Sigma}_E\qu$
and $y_2=\cdots=y_{n_T}=0$. Since $x_1 > y_1$ and the function
$g(z_1,\cdots,z_{n_T})$ defined in (\ref{func_g})
is strictly increasing with respect to $z_1$, we get that $C_s(\qQ)>0$. This completes the proof.

\section{Proof of Lemma \ref{CondPosRateCSI}}\label{ProofLemCondPosRateCSI}

From (\ref{ergoSecRateCSI}) and (\ref{hE}), we get
\al{
C_s(\qQ)&=\log (1+\rho\qh_R^\dagger\qQ\qh_R)\nnb\\
&\quad -\mathbb{E}_{\qh_w}\{\log (1+\rho\qh_w^\dagger{\bs \Sigma}_E^{1/2}\qQ{\bs \Sigma}_E^{1/2}\qh_w)\}.
}
It follows from the Jensen's inequality \cite[p. 25]{Cover} that $\log \mathbb{E}(x) \ge \mathbb{E} (\log x)$.
With this and the fact that
$\mathbb{E}_{\qh_w}\{\qh_w^\dagger \qA\qh_w\}=\mathbb{E}_{\qh_w}\{\mathrm{Tr}(\qA\qh_w\qh_w^\dagger)\}=\mathrm{Tr}(\qA)$, we get
\equ{
C_s(\qQ)\ge \log (1+\rho\qh_R^\dagger\qQ\qh_R)-\log (1+\rho\mathrm{Tr}(\qQ{\bs \Sigma}_E)).
}
Note that $\qh_R^\dagger\qQ\qh_R - \mathrm{Tr}(\qQ{\bs \Sigma}_E) = \mathrm{Tr}(\qQ(\qh_R\qh_R^\dagger - {\bs \Sigma}_E))$.
Let $\qu$ be the eigenvector associated with the largest eigenvalue $\lambda$ of $\qh_R\qh_R^\dagger - {\bs \Sigma}_E$.
Since $\lambda > 0$, we know that $\mathrm{Tr}(\qu\qu^\dagger(\qh_R\qh_R^\dagger - {\bs \Sigma}_E))
=\qu^\dagger(\qh_R\qh_R^\dagger - {\bs \Sigma}_E)\qu=\lambda>0$.
Thus, $C_s(\qQ)>0$ holds for $\qQ=\qu\qu^\dagger$. This completes the proof.

\section{Proof of Lemma \ref{Lem:ExpectCal}}\label{ProofLem:ExpectCal}

From \cite[Eq. (64)]{Moustakas}, we know
\al{
&\mathbb{E}_{\qz}\{\log (1+\rho\qz^\dagger\qQ\qz)\}\nnb\\
=\ &\int_0^\infty \frac{e^{-t}}{t}
\left(1-\frac{1}{\det(\qI+t\rho\qR^{1/2}\qQ\qR^{1/2})}\right)
\mathrm{d}t.\label{Expe_z}
}
By inserting $\qR^{1/2}\qQ\qR^{1/2}=\qU_1\qD_1\qU_1^\dagger$ into (\ref{Expe_z}), we get
\equ{
\mathbb{E}_{\qz}\{\log (1+\rho\qz^\dagger\qQ\qz)\}
=\int_0^\infty e^{-t}
\bigg(\frac{1}{t}-\frac{1}{t\prod_{i=1}^M (1+t\rho d_i)}\bigg)
\mathrm{d}t.\label{ErgodicSecRate}
}
Performing  partial fraction expansion, i.e.,
\equ{
\frac{1}{t}-\frac{1}{t\prod_{i=1}^M (1+t\rho d_i)}
=\sum_{j=1}^M\frac{\rho d_j}{\prod_{i\ne j}^M(1-d_i/d_j)}\frac{e^{-t}}{1+t\rho d_j}
}
and using
\equ{
\int_0^\infty \frac{e^{-t}}{1+ta}\mathrm{d}t=\frac{1}{a}F_1(a),\label{int1}
}
we get (\ref{E_z}). This completes the proof.

\section{Proof of Theorem \ref{theo:1}}\label{proofTheo1}

It follows from (\ref{KKTcond2}) that ${\bs \Psi}\qQ=\qQ{\bs\Psi}=0$, that is, ${\bs\Psi}$ and $\qQ$ commute and
have the same eigenvectors \cite[p.239]{Davis} and their eigenvalue patterns are complementary
in the sense that if $\lambda_i(\qQ) > 0$, then $\lambda_i({\bs \Psi}) = 0$, and vice
versa \cite{Vu}.
This result, when combined with (\ref{KKTcond1}), implies that ${\bs\Theta}$ and $\qQ$
commute and have the same eigenvectors.
Further, we get ${\bs\Theta}\qQ=\qQ{\bs \Theta}=\theta \qQ$, which, when combined with $\mathrm{Tr}(\qQ)=1$ and
the fact $\mathrm{Tr}(\qQ{\bs \Theta})$ is always real, leads to
$\theta=\mathrm{Tr}(\qQ{\bs \Theta})$ and (\ref{basicEq1}) (also see \cite{Li}).

The condition (\ref{basicEq1}) reveals that for the  optimal $\qQ$,  $\qQ{\bs \Theta}$ is a scaled version  of $\qQ$.
Further, the eigenvalues of ${\bs\Theta}$ corresponding to the positive eigenvalues of $\qQ$ are all equal to
 $\mathrm{Tr}(\qQ{\bs \Theta})$, while the remaining eigenvalues of ${\bs\Theta}$ are all less than  or equal
to $\mathrm{Tr}(\qQ{\bs \Theta})$, which follows from (\ref{KKTcond1}), (\ref{basicEq1}) and ${\bs\Psi}\succeq 0$.
Based on the above (\ref{basicEq2}) follows.

\section{Proof of Lemma \ref{Lem:ThetaCalc}}\label{ProofLem:ThetaCalc}

Denote the expectation in the left hand side of (\ref{ExpectTheta}) as $I$.
We write $\qz=\qR^{1/2}\qh_w$ where $\qh_w=[h_{w,1},\cdots,h_{w,n_T}]^T\sim \mathcal{CN}(\mb{0}, \qI_{n_T})$,
and $h_{w,k}$'s
follow i.i.d. $\mathcal{CN}(0,1)$. With this, we have
\equ{
I=\mathbb{E}_{\qh_w}\bigg\{\frac{\rho\qR^{1/2}\qh_w\qh_w^\dagger\qR^{1/2}}{1+\rho\qh_w^\dagger\qR^{1/2}\qQ\qR^{1/2}\qh_w}\bigg\}.
\label{Expect_z_Theta}
}
By inserting $\qR^{1/2}\qQ\qR^{1/2}=\qU_1\qD_1\qU_1^\dagger$ into (\ref{Expect_z_Theta}), we have
\equ{
I=\rho\qR^{1/2}\mathbb{E}_{\qh_w}
\bigg\{\frac{\qh_w\qh_w^\dagger}{1+\rho\qh_w^\dagger\qU_1\qD_1\qU_1^\dagger\qh_w}\bigg\}\qR^{1/2}.
}
Then we use the fact that $\qU\qh_w$ and $\qh_w$ have the identical distributions for any unitary matrix $\qU$
to obtain
\equ{
I=\rho\qR^{1/2}\qU_1\qY\qU_1^\dagger\qR^{1/2}
}
where
\equ{
\qY=\mathbb{E}_{\qh_w}
\bigg\{\frac{\qh_w\qh_w^\dagger}{1+\rho\qh_w^\dagger\qD_1\qh_w}\bigg\}
}
with $(i,j)$th entries given by
\equ{
Y_{ij}=\mathbb{E}_{\qh_w}
\bigg\{\frac{h_{w,i}h_{w,j}^\ast}{1+\rho\sum_{k=1}^M d_k|h_{w,k}|^2}\bigg\}.
}
From the gamma integral \cite{Furrer} we have
\equ{
\frac{1}{a^z}=\frac{1}{\Gamma(z)}\int_0^\infty t^{z-1}e^{-ta}\mr{d}t, \ \mbox{Re}(z)>0, a>0
}
where $\Gamma(z)=\int_0^\infty u^{z-1}e^{-u}\mr{d}u$, we let $z=1$ to obtain
$\frac{1}{a}=\int_0^\infty e^{-ta}\mathrm{d}t$. With this identity, we can write
\equ{
Y_{ij}=\int_0^\infty e^{-t}\mathbb{E}_{\qh_w}
\bigg\{h_{w,i}h_{w,j}^\ast \prod_{k=1}^M e^{-t\rho d_k|h_{w,k}|^2}\bigg\}\mathrm{d}t.\label{Yij}
}
Since
$h_{w,k}$'s
follow i.i.d. $\mathcal{CN}(0,1)$,
we know $Y_{ij}=0$ for $i\ne j$, i.e., $\qY$
is a diagonal matrix with $(k,k)$th entries given by
\al{
Y_{kk}&= \int_0^\infty \frac{e^{-t}} {\prod_{i=1}^M (1+t\rho d_i)}\frac{1}{1+t\rho d_k}
\mathrm{d}t, k\le M\label{IntYkk1}\\
Y_{kk}&=\int_0^\infty \frac{e^{-t}} {\prod_{i=1}^M (1+t\rho d_i)}
\mathrm{d}t, k>M.\label{IntYkk2}
}
These integrals can be easily calculated. Performing  partial fraction expansion
\equ{
\frac{1}{\prod_{i=1}^M (1+t\rho d_i)}
=\sum_{j=1}^M\frac{1}{\prod_{i\ne j}^M(1-d_i/d_j)}\frac{1}{1+t\rho d_j}
}
and using (\ref{int1}) and
\al{
\int_0^\infty \frac{e^{-t}}{(1+ta)(1+tb)}\mathrm{d}t&=\frac{F_1(a)-F_1(b)}{a-b}\label{int2}\\
\int_0^\infty \frac{e^{-t}}{(1+ta)^2}\mathrm{d}t&=F_2(a),\label{int3}
}
we get (\ref{ExpectTheta}). This completes the proof.

\section{Proof of Lemma \ref{Lem:phi_zConvex}}\label{ProofLem:phi_zConvex}

We need to prove that
\equ{
\phi(t z_1 + (1-t)z_2) \le t\phi(z_1) + (1-t)\phi(z_2), \ \forall t\in [0, 1].\label{Convex}
}

Let $\qX_1$ and $\qX_2$ be the optimal $\qX$ associated with $z_1$ and $z_2$.
Consider the problem associated with $t z_1 + (1-t)z_2$, i.e.,
\al{
&\min_{\qX} \ \mathrm{Tr}({\bs \Sigma}_E\qX)\label{SDP}\\
          &\mathrm{s.t.}\quad \mathrm{Tr}(\qh_R\qh_R^\dagger\qX)=(t z_1 + (1-t)z_2)\|\qh_R\|^2, \nnb\\
          &\quad\quad\ \mathrm{Tr}(\qX) =1,\nnb\\
          &\quad\quad\ \qX \succeq 0.\nnb
}
It is easy to verify that $t\qX_1+(1-t)\qX_2$ satisfies the constraints in the problem of (\ref{SDP})
with the corresponding objective value $t\phi(z_1) + (1-t)\phi(z_2)$.
Thus, (\ref{Convex}) holds and $\phi(z)$ is a convex function.

\section{Proof of Lemma \ref{Lem:posiEig}}\label{ProofLem:posiEig}

Let $\lambda>0$ be an eigenvalue of $\qa\qa^\dagger-\qA$, and we have
\equ{
\det(\qa\qa^\dagger-\qA-\lambda\qI)=0.\label{EigEqu}
}
Note that $\qA+\lambda\qI$ is positive definite.
By using the identity $\det(\qB-\qa\qa^\dagger)=(1-\qa^\dagger\qB^{-1}\qa)\det(\qB)$ for a positive definite matrix $\qB$,
it follows from (\ref{EigEqu}) that
\equ{
1-\qa^\dagger(\qA+\lambda\qI)^{-1}\qa=0.\label{EigEqu2}
}
Denote $\ell(\lambda)\triangleq 1-\qa^\dagger(\qA+\lambda\qI)^{-1}\qa$.
It is easy to verify that $\ell(\lambda)$ is a strictly increasing function.
Thus, $\ell(\lambda)$ has only one positive root, and $\ell(0)\ne 0$, i.e., $0$ is not
a eigenvalue of $\qa\qa^\dagger-\qA$. Thus, all other eigenvalues are negative. This completes the proof.

\section{Proof of Lemma \ref{Lem:Expecta}}\label{proofLem:Expecta}

We have
\equ{
\log x > \mathbb{E}(\log Y) \Longrightarrow \mathbb{E} \left( \log \frac{x}{Y}\right)>0.
}
By using Jensen's inequality \cite[p. 25]{Cover}, we have
\al{
&\log \mathbb{E}\left(\frac{x}{Y}\right)\ge \mathbb{E}\left(\log\frac{x}{Y}\right)\nnb\\
\Longrightarrow \ & \mathbb{E}\left(\frac{x}{Y}\right) > 1 \nnb\\
\Longrightarrow \ & x \mathbb{E}\left(\frac{1}{Y}\right) > 1 \nnb\\
\Longrightarrow \ & \frac{1}{x} < \mathbb{E}\left(\frac{1}{Y}\right).
}

%================================================ Statistical info

% that's all folks
\end{document}